\documentclass[%
 aip,
 amsmath,amssymb,
preprint,%
]{revtex4-1}

\usepackage{graphicx}
\usepackage{dcolumn}
\usepackage{bm}

\usepackage[utf8]{inputenc}
\usepackage[T1]{fontenc}
\usepackage{mathptmx}

\begin{document}

\preprint{AIP/123-QED}

\title[A Semi-Grand Canonical Kinetic Monte Carlo study of Single-Walled Carbon Nanotubes growth]{A Semi-Grand Canonical Kinetic Monte Carlo study of Single-Walled Carbon Nanotubes growth}

\author{Georg Daniel Förster}
\affiliation{Aix Marseille Univ - CNRS, CINaM UMR 7325, Campus de Luminy - Case 913, 13288 Marseille Cedex 09, France.}

\author{Thomas D. Swinburne}
\affiliation{Aix Marseille Univ - CNRS, CINaM UMR 7325, Campus de Luminy - Case 913, 13288 Marseille Cedex 09, France.}

\author{Hua Jiang}
\affiliation{Department of Applied Physics, Aalto University School of Science, P.O. Box 15100, FI-00076 Aalto, Finland.}

\author{Esko Kauppinen}
\affiliation{Department of Applied Physics, Aalto University School of Science, P.O. Box 15100, FI-00076 Aalto, Finland.}

\author{Christophe Bichara}
\affiliation{Aix Marseille Univ - CNRS, CINaM UMR 7325, Campus de Luminy -- Case 913, 13288 Marseille Cedex 09, France.}
\email{xtof@cinam.univ-mrs.fr}

\date{\today}

\begin{abstract}

Single-walled carbon nanotubes exist in a variety of different geometries, so-called chiralities, that define their electronic properties. Chiral selectivity has been reported in catalytic chemical vapor deposition synthesis experiments, but the underlying mechanisms remain poorly understood. In this contribution, we establish a simple model for the prediction of the growth rates of carbon nanotubes of different chiralities, as a function of energies characterizing the carbon nanotube-catalyst interface and of parameters of the synthesis. The model is sampled efficiently using kinetic Monte Carlo simulations in the semi-grand canonical ensemble, uncovering the interplay of the external experimental conditions and the configuration and energetics of the interface with the catalyst. In particular, the distribution of chiral angle dependent growth rates follows non-monotonic trends as a function of interface energies. We analyze this behavior and use it to identify conditions that lead to high selectivity for a variety of chiral angles.

\end{abstract}

\maketitle

\section{Introduction}

Single-Walled Carbon Nanotubes (SWNTs) are crystalline structures that can be seen as rolled up graphene sheets with a remarkably large aspect ratio, reaching up to tens of centimeters in length and a few nanometers in diameter. Since their first observation and characterization~\cite{Iijima93} tremendous research efforts went into the development of efficient synthesis techniques, the most prominent being the catalytic chemical vapor deposition (CVD), in which a carbon-rich gaseous feedstock (e.g., CO, CH$_4$, ethanol) is being decomposed at high temperature, on the surface of a catalyst, often a transition metal, to nucleate and grow a tube~\cite{Jourdain13}. A number of groups~\cite{Sundaram2011, Yang2014, Zhang2017, Wang2018} have reported the selective growth of tubes with a particular helicity (or chirality) characterized by the so-called Hamada indices $(n, m)$. However, we are far from understanding all aspects of this control of chirality which, if routinely achieved, would promote the development of technologies based on the exceptional electronic properties of SWNTs~\cite{Rao2018}. A reason for the difficulties with selective growth is that SWNTs may appear in a (theoretically) infinite number of polymorphs with very small energy differences between them, pushing the crystal growth science to its limits. \\

While SWNTs are indeed crystals, they display characteristic properties making them different from 3-dimensional materials. The first obvious difference is that the active area for carbon incorporation in the tube during CVD growth is made of only a few tens of atoms at the interface between the tube and its seeding catalyst. Because of the high temperature of the CVD synthesis and of its nanometric size,  this interface is experimentally difficult to characterize and even more to control.  A second aspect is that some local topological defects, such as a pair of non-hexagonal carbon rings, have the potential to change the crystal structure of the tube~\cite{Lambin1999}. Point defects are present and thermodynamically stable in 3-dimensional materials without altering their crystallographic structure. However, this is not always the case in the present instance, and the fact that perfect SWNTs with a defined structure can be grown up to centimeter length~\cite{Zhang2017a, Bai2020}, implies a negligible concentration of topologically active defects in this size range. 

Classically, crystals are formed by nucleation and growth. The former is dominated by the thermodynamic stability of the nucleus, while kinetics is central to the latter. We recently developed a thermodynamic model of the catalyst-tube interface where the fluctuations of the interface structure and the associated entropy were found to be responsible for the stability of the so-called "chiral" tubes, those which are neither armchair nor zigzag~\cite{Magnin2018}. This thermodynamic model  allowed to identify the most stable nanotube structure as a function of three parameters: the interface energies of the armchair or zigzag edge atoms and the temperature. Thus, distributions of the probability of nucleation of the tubes as a function of their chirality could be established, which compare favorably to the experimental results, but have a tendency to be too broad, probably because the growth kinetics were not taken into account.

Our goal here is thus to explore growth kinetics, using the simplest possible kinetic Monte Carlo (kMC) simulations on a fixed lattice unique to each helicity. A similar route has recently been proposed for studying the growth and etching of graphene flakes on catalytic substrates~\cite{Kong2021}. In our approach which is a development of the work by Dumlich and Reich~\cite{Dumlich2010} using more powerful computer simulation tools, we consider the growth from the point of view of the tube, with the underlying idea that the interplay between different edge geometries that are constrained by the helicity of the tube and the energies of the interface in contact with the catalyst, could lead to a chirality-selective growth. In this approach, the thermochemistry of the catalytic decomposition of the gaseous precursor, the diffusion of carbon atoms to the nanotube lip, and the amount of carbon dissolved in the catalytic nanoparticle (NP) that controls the growth mode~\cite{He2018}, are considered as elements of a global process, independent of the tube's helicity, that delivers carbon atoms at a given chemical potential to the tube edge and thus drives the growth or etching of the tube.

\begin{figure*}[htb!]
\begin{center}
\includegraphics{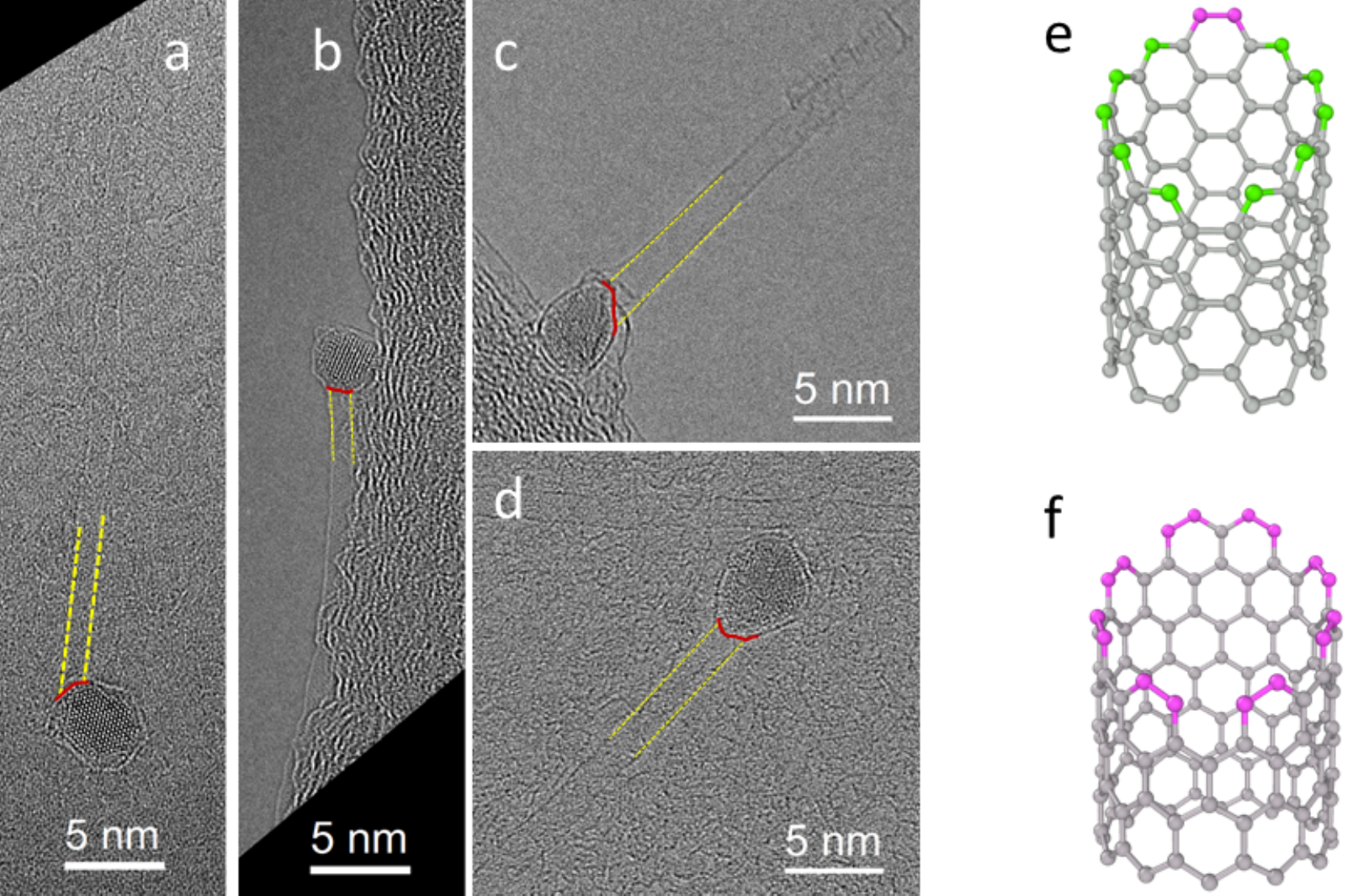}
\caption{\label{fig:Fig1}(Color online) Different carbon nanotube--catalyst interfaces: (a-d) HRTEM images of carbon nanotubes (yellow) attached to catalyst nanoparticles. The interfaces are highlighted in red. In some cases the center of the SWNT is not aligned with the catalyst which leads to oblique interfaces. (e) Armchair tubes, here $(6,6)$, can have an oblique rim of $n+m=12$ atoms with up to $2n-2=10$ zigzag (green) and $2$ armchair (pink) atoms. (f) An oblique interface of the zigzag $(12,0)$ tube has more than $n+m=12$ atoms, here $16$.}
\end{center}
\end{figure*}

In the very simple thermodynamic model we proposed, the interface of a $(n,m)$ tube was assumed to be perpendicular to the tube axis, with $n+m$ carbon atoms in contact with the catalyst, $2m$ of them in armchair configuration and the remaining $n-m$ in zigzag configuration. Such interfaces are indeed observed~\cite{Magnin2018}, but recent High Resolution Transmission Electron Microscopy (HRTEM) observations of tube--catalyst interfaces after the synthesis show some variability of the interface structure. Nanotubes were synthesized using the floating catalyst CVD method~\cite{Liao2018}. HRTEM samples were collected directly from the CVD reactor by placing the TEM grid (copper) with thin holey carbon supporting films (Agar Scientific Ltd.) on a filter for 20 seconds. A double aberration-corrected JEOL-2200FS microscope (JEOL Ltd.) was employed for high-resolution observation operated at an acceleration voltage of 200kV. In order to enhance the contrast of carbon nanotubes sitting on thin carbon films, HRTEM images were taken at slightly over-focussed conditions. A number of NPs were found isolated, either inactive or detached from the tube, but a few tens of tube--catalyst interfaces could be observed, showing that nanotubes may display more complex, oblique interfaces. Typical examples are displayed in Fig.~\ref{fig:Fig1}. However, it should be noted that the structure of the interface might be modified during the cooling process, but, using a similar approach, Fiawoo \textit{et al.} could identify different growth modes~\cite{Fiawoo2012} that have been validated since then~\cite{He2018}. Such interfaces often correspond to the tube axis being off the center of the NP. This might help quantifying how oblique interfaces are. In the present example, a majority of the observed interfaces were oblique, but this probably depends on the CVD conditions, and a more detailed study should be performed.\

This points to an important observation, overlooked in Ref.~\onlinecite{Magnin2018}: if the number of carbon atoms at the interface is kept equal to $n+m$, there is only one possible edge for zigzag tubes, while tubes with larger chiral angles $\theta$ can exhibit a large number of edges, including more oblique ones. The most oblique interface of a $(n,n)$ tube contains $2n-2$ zigzag and $2$ armchair atoms. This asymmetry has consequences on both thermodynamic stability and growth mechanisms. \

This article is organised as follows. We start by defining our lattice model and the associated energies. We then present the kMC algorithm. Since its application in a grand canonical ensemble is not so common, we demonstrate the validity of our approach that makes it possible to simulate growth and etching of tubes. This enables us to calculate growth rates for a large set of $96$ tubes with different chiralities, that depend on the five parameters of the energy model and the growth conditions (carbon chemical potential and temperature). We then try to disentangle the influence of each parameter and finally display three typical situations where a high selectivity based on large differences of growth rates can be expected.

\section{The model}
We model the CNT growth process using a lattice gas, where lattice sites are either empty or occupied by a carbon atom. Carbon atoms in the lattice have either three carbon neighbors when they are within the tube, forming ``bulk'' sites, denoted $C_3$, or two carbon neighbors and one empty neighbor site, forming either zigzag $C_Z$ or armchair $C_A$ edge sites. The former have two neighboring $C_3$ sites, the latter have one $C_3$ and one two-fold coordinated sites. We also consider C atoms with only one C neighbor, attached to either $C_A$ or $C_Z$ edge sites, denoted $C_A^1$ and $C_Z^1$, respectively. These high energy configurations approximate transition states for the addition and removal of atoms during the kMC simulations, as discussed below. Isolated C atoms with no C neighbor and other types of two-coordinated atoms, forming chains, are ruled out. Thus there are $N_A$, $N_Z$, $N_A^1$, $N_Z^1$ undercoordinated atoms of type $C_A$, $C_Z$, $C_A^1$, $C_Z^1$ in contact with a catalyst. The energies $E_A$, $E_Z$, $E_A^1$, $E_Z^1$ of these states are interface energies, as opposed to those of atoms at an open nanotube edge without contact to a catalyst. \

\begin{figure}[htb!]
\begin{center}
\includegraphics{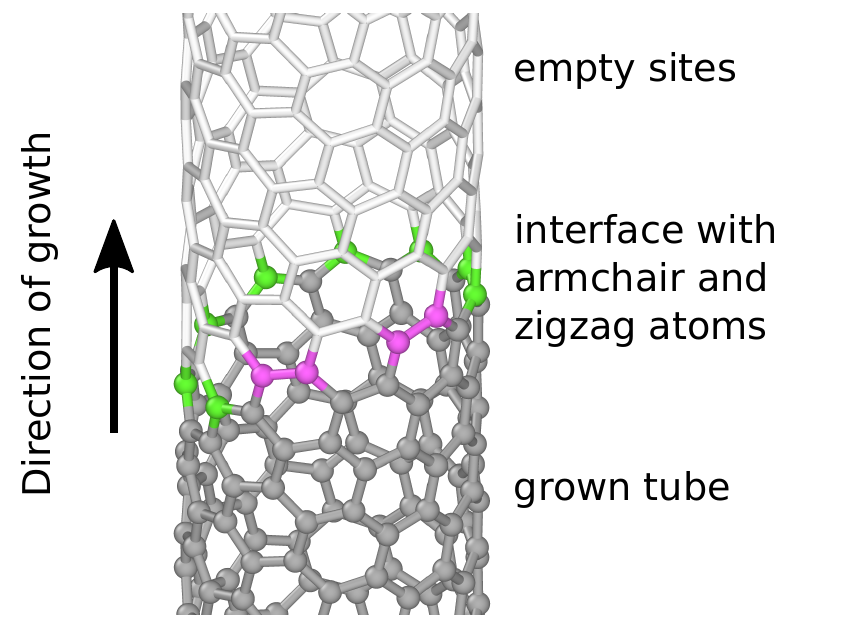}
\caption{\label{fig:Fig2}(Color online) Schema for the lattice kMC simulations of the growth of carbon nanotubes. Carbon atoms in zigzag and armchair configuration in contact with the catalyst (not represented) at the edge of the tube are shown as green and pink balls, respectively. Atoms in the bulk part of the tube are represented as gray balls and the empty part of the lattice is drawn as white sticks.}
\end{center}
\end{figure}

Such interface energies are very difficult to measure experimentally, but have been evaluated using DFT calculations for different catalysts. Most of these calculations were performed on small systems relaxed at zero temperature, and rely on assumptions regarding the structure of the catalyst NP and its surface. Catalysts probed in Ref.~\onlinecite{Ding2008} (Au, Cu, Pd, Ni, Co, Fe) show a stronger interaction with zigzag tubes, but an investigation of a larger set of metals~\cite{Hedman2019} indicates that this is not a general trend. Interestingly, using DFT-based molecular dynamics at high temperature for very small bimetallic clusters, Qiu and Ding~\cite{Qiu2019} showed that the nanotube--metal interaction can be strong enough to drag the metal with the higher carbon affinity to the lip of tube, thus modifying the structure of the nanoparticle. Recent theoretical investigations~\cite{Penev2018, Bets2019} suggest that ordering or phase separation tendencies between armchair and zigzag species at the edge of the tube can be induced by the catalyst. We thus include an ordering energy term $E_{A/Z} = \bm{r} \: \varepsilon_{A/Z}$, where $\bm{r}$ is the number of contacts between armchair and zigzag species and $\varepsilon_{A/Z}$ the energy per contact, to account for these tendencies. As illustrated in Fig.~\ref{fig:Fig2}, the total energy of a configuration is the sum of two terms: a ``bulk'' term that depends on the energy and number of $C_3$ atoms and an interface energy term $E_{\text{Int}}$ that includes carbon atoms with less than three carbon neighbors. Setting the energy reference to zero for $C_3$ carbon atoms in the lattice, we end up with a model in which the total energy is reduced to the interface energy that writes:
\[
E_{\text{Int}} = N_A\:E_A + N_Z\:E_Z + N_A^1\:E_A^1 + N_Z^1\:E_Z^1 + \bm{r} \: \varepsilon_{A/Z}
\]
Note that all interface energy terms for 2- and 1-fold coordinated C atoms are positive, while a negative (positive) $\varepsilon_{A/Z}$ favors armchair--zigzag alternation (separation) at the edge. The typical range of $E_A$ and $E_Z$ is $0.0 - 0.5$ eV/atom~\cite{Magnin2018}, while calculations of the interface energies of various SWNTs in contact with the W-terminated (0 0 3) surface of a Co$_7$W$_6$ alloy~\cite{Penev2018, Bets2019} led to positive values of $\varepsilon_{A/Z}$, that would correspond to about $+0.30$~eV in our model. We note that none of these calculations dealt with the real conditions encountered during a CVD synthesis, in particular concerning the NP structural stability~\cite{An2019} and the role of carbon dissolution in the catalyst~\cite{Yang2019}. In the present approach, we consider these interface energies as parameters, allowing them a large flexibility to probe the limits of the model.

During the course of the kMC simulations, carbon atoms are incorporated in or removed from the lattice. The corresponding statistical mechanical framework is the so-called semi-grand canonical ensemble (sGC-kMC): the total number of occupied lattice sites fluctuates in a manner controlled by the temperature $T$ and a chemical potential difference $\Delta\mu_C$ between empty and occupied sites in the lattice. Defining $\mu_C^{\text{tube}}$ as the chemical potential of a carbon atom incorporated in the tube and $\mu_C^{\text{cat}}$ as that of a C atom in or on the catalyst, defined by the thermochemistry of the decomposition of the precursor, we simulate the CVD process by identifying $\Delta\mu_C$ as the difference $\mu_C^{\text{cat}} - \mu_C^{\text{tube}}$. 

The growth and etching mechanisms are driven by $\Delta\mu_C$. Equilibrium corresponds to $\Delta\mu_C=0$. When $\Delta\mu_C\neq0$, the system is driven out of equilibrium with unbounded growth for $\Delta\mu_C>0$ and etching for $\Delta\mu_C<0$. We note that the free energy landscape of our simple model is closely related to the discrete state Brownian motor, a classic model of non-equilibrium statistical mechanics\cite{Reimann2002}. \

\section{\label{sec:kmc}Kinetic Monte Carlo algorithm}

We wish to understand how the growth rate for the chirality $(n, m)$ is influenced by the energy parameters defined above and the CVD conditions ($T$, $\Delta\mu_C$) by sampling the structure of the tube/catalyst interface in Monte Carlo runs. Furthermore, our goal is to identify conditions of high selectivity for certain chiralities. We thus have to extend the classical kMC algorithm~\cite{Voter} to grand canonical conditions. This has already been done~\cite{Ustinov2012, Tan2016} in the context of fluid systems. Barriers correspond to rare events, that are sampled efficiently in kMC, giving thus access to the physically important phenomena. In the following, performing a set of kMC simulations under the same conditions, we compute growth rates of 96 SWNTs with chiralities ranging from $(5, 4)$ to $(18, 1)$ (diameter range: 6.0--14.8\AA). KMC approaches have already been used in the context of SWNT growth, in particular in an early work by Maiti \textit{et al.}~\cite{Maiti1995}, and more recently by Li \textit{et al.}~\cite{Li2015}. The main differences in the present work are that we use very simple lattice systems to enable extremely fast simulations, which in addition allow the addition and removal of carbon atoms to exhibit both growth and etching.

A key challenge when building kMC models is the identification of all possible atomic level mechanisms, in general a highly non-trivial task\cite{Swinburne2018}. In the present case the actual atomic scale growth process is extremely complex, dependent on the nature of the carbon precursor, the catalyst, and the specific CVD parameters. Using Density Functional Theory (DFT) calculations, Shibuta \textit{et al.}~\cite{Shibuta2013} showed that the decomposition of a simple methane molecule on a Ni surface involves multiple steps and a subsurface burying of the dissociated carbon atoms. Decomposition of the different bonds of the more complex ethanol molecule has been shown to depend on the surface sites of either Fe, Co or FeCo catalysts~\cite{Fukuhara2019}. Closer to the present context, Penev \textit{et al.}~\cite{Penev2018} showed how carbon dimers can be attached at an armchair-zigzag contact, forming a pentagon/heptagon pair which can lead to a tube chirality change if it persists. \

\begin{figure}[htb!]
\begin{center}
\includegraphics{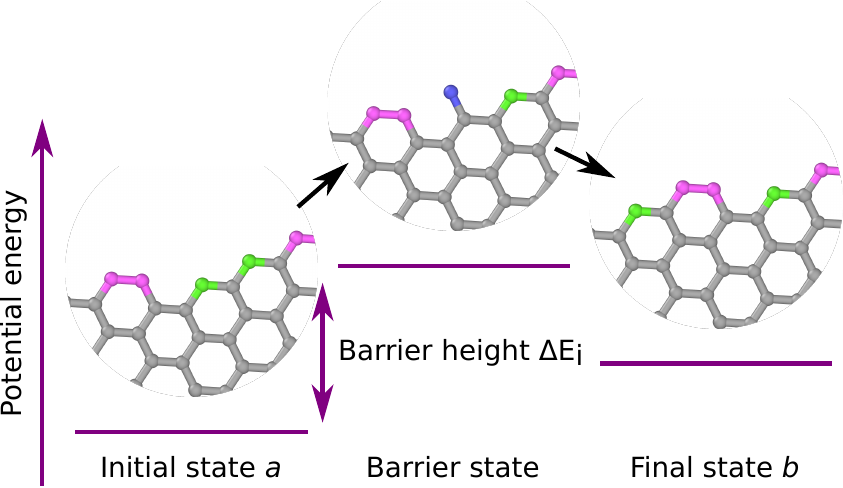}
\caption{\label{fig:Fig3}(Color online) Basins (initial and final state) and barrier states encountered in the kMC algorithm. The barrier states always contain a mono-coordinated atom (represented as a blue sphere) at the interface with the catalyst, while the basin states contain only zigzag and armchair atoms (green and violet spheres, respectively). Energy parameters are chosen in such a way that only positive effective barrier heights are encountered in the simulations.}
\end{center}
\end{figure}

Such complex mechanisms cannot be taken into account in our fixed lattice simulations; we retain only a phenomenological free energy barrier for the addition or removal of two carbon atoms at selected edge sites. In the same spirit as the algorithm described by Voter~\cite{Voter} for the canonical ensemble, during our sGC-kMC simulations, moves are randomly selected from a set of insertion and removal events, transitioning from a basin state $a$ to a basin state $b$. Carbon atoms can be deposed on any $C_A$ or $C_Z$ site that allows to complete a six-membered carbon ring upon addition of a second atom, as illustrated in Fig.~\ref{fig:Fig3}. As the added carbon atom is a high energy, weakly metastable state, we assume that in a period of time much shorter than the typical residence time in any basin state, this atom either detaches or is stabilized through the addition of a second carbon atom. Whilst this is consistent with the degree of approximation inherent in our simplified model, we note that a refined or even completely distinct mechanism would only change the value of the free energy barrier for this process, not the catalogue of available moves in our model. As the transition state contains one additional atom, we assume that the free energy of this state changes by $-\Delta\mu_C$, which is expected to be valid provided that the model transition state has a free energy greater than the initial and final states.


At each kMC step an exhaustive list of such insertion and removal events is established and the energy of each of the corresponding barrier states is evaluated. These rules imply that for the basin states all $n+m$ edge atoms are either of type $C_A$ or $C_Z$. The edge configurations of the barrier states contain $n+m-1$ atoms of these types and one mono-coordinated atom of type $C_A^1$ or $C_Z^1$. Each full Monte Carlo step, i.e. from one basin state to the next, fills or empties two lattice sites, thereby creating or eliminating two $C_3$ atoms, respectively. We identify 14 possible barrier states, 5 for insertion and 9 for removal, as shown in table~\ref{tab:barriers}. A graphical representation of these barrier states can be found in Fig.~\ref{fig:barriers}. Depending on nanotube chirality and model parameters not all of them necessarily occur in the simulations. In particular, in the case of $m=0$, there are no barrier states, and therefore these tubes cannot grow within this model. In the case of $m=1$, only a subset of 4 barriers are available, due to geometric constraints.

\begin{table}
\caption{\label{tab:barriers}List of energy barriers occurring in the kMC simulations given as a function of the difference in the number of geometric features of the interface.}
\begin{ruledtabular}
\begin{tabular}{c c c c c c c}
Type & ID & $ \Delta N^{1}_{A}$ & $ \Delta N^{1}_{Z}$ & $ \Delta N_{A}$ & $ \Delta N_{Z}$ & $ \Delta \bm{r}$ \\ \hline
insertion & 1 & 0 & 1 & 0 & -1 & 0  \\
          & 2 & 1 & 0 & -2 & 1 & 2  \\
          & 3 & 0 & 1 & 0 & -1 & -2 \\
          & 4 & 1 & 0 & -2 & 1 & -2 \\
          & 5 & 1 & 0 & -2 & 1 & 0  \\ \hline
removal   & 6 & 0 & 1 & -2 & 1 & -2 \\
          & 7 & 1 & 0 & 0 & -1 & -2 \\
          & 8 & 0 & 1 & -2 & 1 & 2  \\
          & 9 & 0 & 1 & 0 & -1 & -2 \\
          & 10 & 1 & 0 & -2 & 1 & 0  \\
          & 11 & 0 & 1 & -2 & 1 & 0  \\
          & 12 & 1 & 0 & 0 & -1 & 0  \\
          & 13 & 1 & 0 & -2 & 1 & -2 \\
          & 14 & 0 & 1 & 0 & -1 & 0  \\
\end{tabular}
\end{ruledtabular}
\end{table}

The energy of the barrier states is calculated from table~\ref{tab:barriers} and we notice that the energy barriers for carbon insertion and removal are in principle asymmetric. We neglect entropic contributions from thermal vibrations, meaning all free energies are determined solely by the energetic model and the chemical potential difference $\Delta\mu_C$.
Growth moves (C insertion) are associated with 5 barriers, with 3 independent parameters only, that are: $(E_Z^1 - E_Z)$, $(E_A^1 - 2 E_A + E_Z)$ and $\varepsilon_{A/Z}$. Etching moves (C removal) correspond to 9 barriers, with 5 independent parameters only, that are the same as above plus $(E_A^1 - E_Z)$ and $(E_Z^1 - 2 E_A + E_Z)$. We use the rejection-free algorithm as detailed by Voter~\cite{Voter}, deriving from the ``n-fold way'' algorithm by Bortz \textit{et al.}~\cite{Bortz1975}: From the full list of available barriers at each given MC step, a barrier $i$ is randomly selected with a weight proportional to the corresponding rate constant, calculated as:
\begin{equation}
k_i=\nu_0 \: \text{exp}\left(-\left(\Delta E_i - \Delta\mu_C \Delta N\right) /k_BT\right)
\label{eq:defRate}
\end{equation}
where $\Delta E_i$ is the energy barrier height, $k_B$ the Boltzmann constant and $\Delta N = +1$ when adding or $-1$ when removing a C atom. The exponential term is simply proportional to the ratio of the thermodynamic probabilities of the barrier and the basin $a$ states, in the semi grand canonical ensemble. The important factor that controls the process is $\left(\Delta E_i - \Delta\mu_C \Delta N\right)$. In the following we refer to it as the ``effective barrier''. In principle, the attempt frequency $\nu_0$ can be estimated from experimental data or molecular dynamics simulations. Since the purpose of this paper is to compare growth rates of CNTs with different chiralities, which will be unaffected by the absolute value of $\nu_0$, we take $\tau=10^6 \nu_0^{-1}$ as unit of time. At each step, the clock is advanced by drawing a random number from the distribution $p\left(t\right) = k_{\text{tot}}\text{exp}\left(-k_{\text{tot}}t\right)$, with $k_{\text{tot}}=\sum_{i}k_i$. This means that, with this rejection-free algorithm, the selected move is systematically accepted, but the probability of selecting a particular barrier depends on its height. The corresponding time increment for a particular move, which defines the kinetics, depends on the height of all barriers.

\subsection{Observables}

Growth rates can be defined either as the mass uptake, or as the tube length added, both scaled per unit of time $\tau$. When dividing the latter by the tube diameter, the two are proportional. The former therefore favors smaller diameter tubes, and the latter larger diameter ones. Except for tubes with $m=0$, which do not grow, and $m=1$ with four available barriers, independent of diameter, the number of available barriers is roughly proportional to the circumference of the tubes. The clock advances proportional to $k_{\text{tot}}^{-1}$, and therefore roughly inversely proportional to the tubes circumference for $m>1$. Therefore, for these tubes, the length of tube added per unit of time is approximately independent of the diameter. As a consequence, significant differences in growth rate with respect to diameter cannot be expected and our calculated growth rates mostly depend on chiral angles. \
In this model, all tubes with $m\geq1$ grow to some extent. The chiral angle of a $(n,m)$ tube is defined by $\theta = \arctan\left(\frac{\sqrt{3}m}{m+2n} \right)$. Values range between $0^\circ$ for zigzag $(n,0)$ and $30^\circ$ for armchair $(n,n)$ tubes. We group the chiral angles of the tubes in $2^\circ$ bins, and average the corresponding growth rates to analyze the selectivity. We define the selectivity $S(\theta)$ as the growth rate in terms of tube length $G(\theta)$ divided by the sum of all growth rates of the set of all 15 bins.
\[
S(\theta) = \frac{G(\theta)}{\sum_{\theta} G(\theta)}
\]

\section{Results}

\subsection{Validation of the modeling}

Before analyzing the effect of the model parameters on growth rates, we establish a baseline result which can be compared to an analytical expression for the growth rate. For this, two-fold coordinated atoms at the interface are assigned $E_A=E_Z=0.15$~eV, one-fold coordinated atoms $E_A^1=E_Z^1=0.8$~eV, and the ordering energy $\varepsilon_{A/Z}=0.0$~eV. In this case, all energy barriers are equal, $\Delta E_i=0.65$~eV. The temperature is set to $T = 1000$~K and the chemical potential difference to $\Delta\mu_C=0.1$~eV. In the following we use these values in simulations of $10^5$ steps, where not indicated otherwise. 

\begin{figure}[htb!]
\begin{center}
\includegraphics{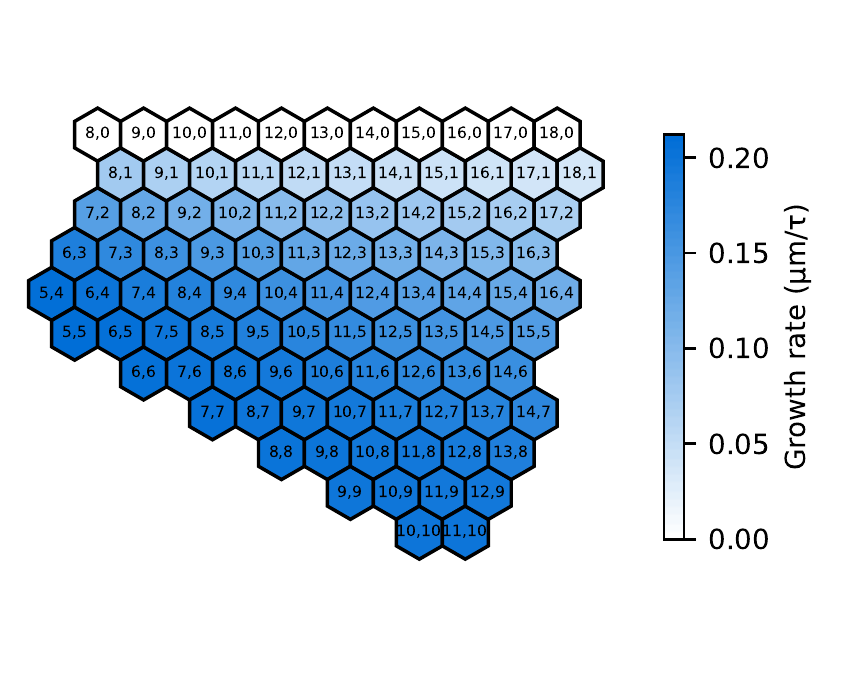}
\caption{\label{fig:Fig4}(Color online) Growth rates for individual chiralities in the ``baseline'' case : $E_A=E_Z=0.15$~eV, $E_A^1=E_Z^1=0.8$~eV, $T=1000$~K and $\Delta\mu_C=0.1$~eV. Zigzag $(n,0)$ tubes do not grow, while growth rates increase with chiral angle and only very weakly depend on the diameter. The animation shows maps for $0.0$~eV~$\leq E_Z \leq$0.5~eV including the corresponding barrier heights (Multimedia view).}
\end{center}
\end{figure}

With these baseline parameters, tubes with large chiral angles are favored, with a relatively low selectivity (see Fig.~\ref{fig:Fig4}). As expected, tube diameter does not affect growth rates very much. More interestingly, because all barriers and hence all time increments are drawn from the same probability distribution (assuming equal $n+m$), the faster growth on the armchair side results from the larger number of available of edge sites allowing the formation of a new hexagon by adding two carbon atoms. Ultimately, because each dimer addition at the edge leads to exchanges of armchair and zigzag edge atoms, this is related to the larger configurational entropy of the edges on the armchair side.

\begin{figure}[htb!]
\begin{center}
\includegraphics{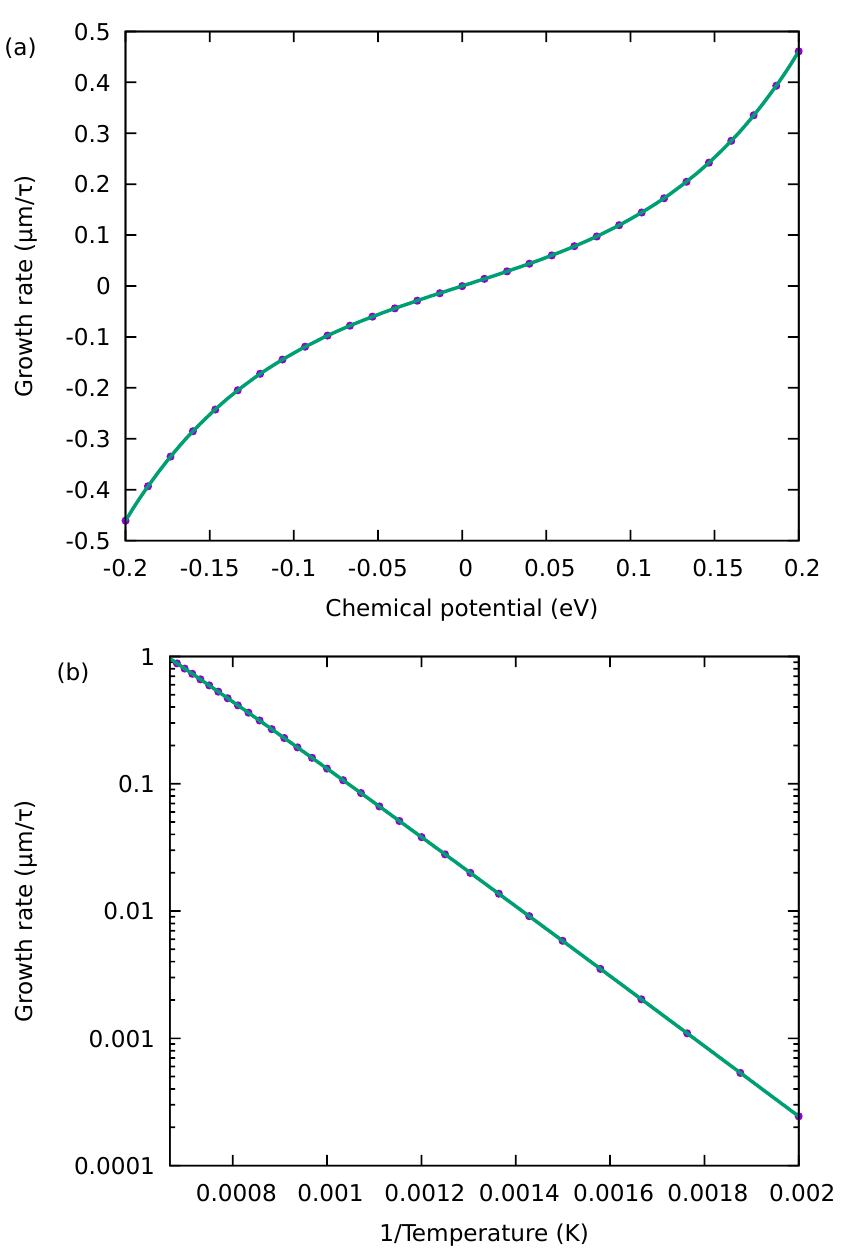}
\caption{\label{fig:mu_temp}(Color online) Effect of chemical potential (a) and temperature (b) on overall growth rates in the case of the baseline parameters ($E_A=E_Z=0.15$~eV, $E_A^1=E_Z^1=0.8$~eV, and $\varepsilon_{A/Z}=0$~eV). Data (purple) and eq.~(\ref{eq:g_tot}) (purple).}
\end{center}
\end{figure}

We then analyzed the effect of the chemical potential on growth rate. Except for zigzag tubes, which are frozen at any value of $\Delta\mu_C$, all tubes are decaying for $\Delta \mu_C < 0$, and growing for $\Delta \mu_C > 0$. At $\Delta \mu_C= 0$, the tube height fluctuates around its initial position, with a constant average value. In the baseline case, the fastest growing tubes are also those that are etched away the fastest, and the ratios of the growth rates for the different chiralities remain constant as the chemical potential is changed. Figure~\ref{fig:mu_temp}(a) shows the average growth rate of the set of 96 CNTs as a function of the chemical potential. For this baseline parameter set it can be shown that (see appendix~\ref{sec:si}),
\begin{equation}
 G_{\text{tot}} \left(T,\; \Delta\mu_C \right) = 4 \times N \nu_0 \: \text{exp}\left(- \Delta E /k_BT\right) \text{sinh}\left( \Delta\mu_C /k_BT\right),
\label{eq:g_tot}
\end{equation}
where we find $N=5.207\pm 0.004$, the average number of insertion and removal sites. 
With varying $\Delta\mu_C$, we confirm that the SWNT height obeys the fluctuation-dissipation theorem as $\Delta\mu_C\to0$, see appendix~\ref{sec:fluctuation}, demonstrating that our simulations are sampling the correct thermodynamic ensemble.

In order to evaluate the effect of temperature on the growth rate, simulations in the temperature range of 500--1500~K were carried out in the case of the baseline parameters with $\Delta\mu_C=0.1$~eV. As with the dependence on $\Delta\mu_C$, the temperature did not affect the ratios of the growth rates of the different chiralities. Fig.~\ref{fig:mu_temp}(b) shows the growth rates averaged over the full set of CNT chiralities as a function of inverse temperature. Adjusting the above expression, $N=5.2062\pm 0.009$ is obtained again.

In this series of simulations, where baseline parameters have been used, neither temperature nor chemical potential significantly affects the proportionality constants between the growth rates of the different chiralities. However, looking at the components of the energy barriers in table~\ref{tab:barriers} and the expression of the rate constant [eq.~(\ref{eq:defRate})], we can see that other sets of parameters may exist where the growth rates are not always proportional. 
In particular, when energy barriers are not equal, growth and etching rates can be asymmetric with respect to chemical potential.

\subsection{Sensitivity of the model parameters}

In the following, we try to understand the influence of the interface energy parameters on the $(n,m)$ dependent growth rates, and to identify conditions that could lead to a chirality-selective growth of SWNTs. As discussed in section~\ref{sec:kmc}, these energies set the heights of the barriers that can be encountered during the simulations. \

\begin{figure}[htb!]
\begin{center}
\includegraphics{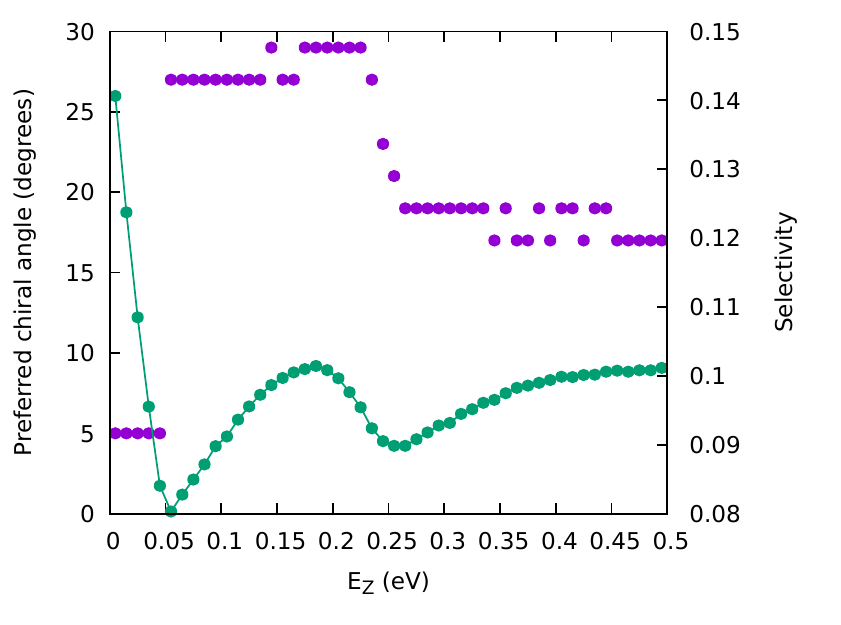}
\caption{\label{fig:zz}(Color online) Favored chiral angle (purple, left y-axis) and growth selectivity (green, right y-axis) as a function of the interface energy of zigzag atoms.}
\end{center}
\end{figure}

\begin{figure}[htb!]
\begin{center}
\includegraphics{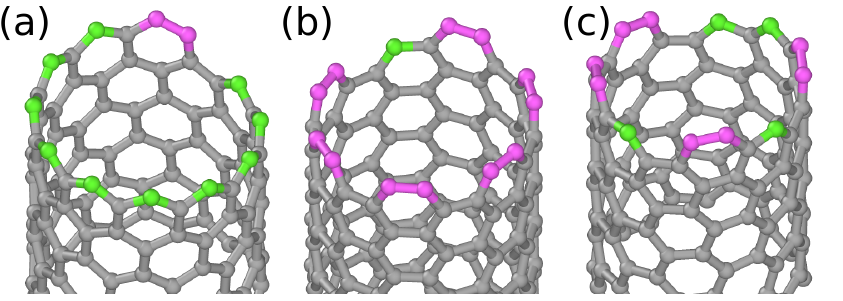}
\caption{\label{fig:growth_modes}(Color online) (a) Spiral growth for a (11, 1) tube at vanishing $E_Z$, (b) Spiral growth for a (7, 6) tube with $E_A=0.05$~eV and $\varepsilon_{A/Z} = 0.15$~eV, and (c) Non-spiral growth of a (8, 4) tube with parameters corresponding to Fig.~\ref{fig:maps}(c). Armchair and zigzag atoms are represented as violet and green spheres respectively. The animations show the growth mechanisms in the three situations, but it is important to note that the sequence of configurations does not include the time delays introduced by the kMC algorithm and hence do not give any hint on the kinetics (Multimedia view).}
\end{center}
\end{figure}

We begin the discussion with the influence of the interface energy $E_Z$ of zigzag atoms, keeping the other energy parameters at their baseline values. The evolution of the growth rates as a function of chirality and $E_Z$ can be appreciated from the animated Fig.~\ref{fig:Fig4}. Results are quantified in Fig.~\ref{fig:zz} that shows the chiral angle of the fastest growing tubes as a function of $E_Z$ and the corresponding selectivity. At vanishing $E_Z$, the growth rate of $(n, 1)$ tubes is the largest and the growth is highly selective, because $(n,1)$ tubes can grow with $N_Z=n+m-2 = n-1$ zigzag edge atoms that are energetically favored, and only $N_A=2$ armchair edge atoms. Moreover, growth is mostly limited to these $(n, 1)$ tubes. As can be seen from Fig.~\ref{fig:growth_modes}(a), the growth proceeds by spiraling around the tube axis which is the only growth process available for tubes with $m=1$. Only insertion barriers 1 and 4, and removal barriers 13 and 14 are utilized. All four barriers are available at every kMC step. However, with the present choice of parameters, barrier 4 has an effective height of only 0.4~eV, compared to the second lowest barrier 13 with 0.6~eV. Therefore barrier 4 is chosen in most of the MC steps. The time increment is small as it is dominated by this low energy barrier, which in turn leads to high growth rates. For other chiralities, the growth is also dominated by barrier 4, which is however, not always available. In these cases, more complex growth patterns emerge that are efficiently sampled in kMC. However, the available barriers are higher and time advances by a large increment, which significantly decreases overall growth rates for any tube with $m > 1$. 

At intermediate values of zigzag energy ($E_Z\approx 0.15$~eV), we recover the baseline condition, where armchair tubes grow the fastest, but with a relatively low selectivity (about a third of what was achieved with the tubes with $m=1$). The transition between favoring tubes with $m=1$ to $n=m$ is abrupt at $E_Z\approx 0.05$~eV. For $E_Z\approx 0.15$~eV, only two different values of effective barrier energies exist: all insertion barriers are at 0.55~eV and all removal barriers at 0.75~eV. The insertion barriers are therefore selected completely randomly. The number of available sites grows with chiral angle, as insertion cannot occur at two neighboring zigzag edges with the same inclination. More available barriers at higher chiral angles directly translates into a smaller time increment which leads in turn to higher growth rates for these chiralities. Increasing $E_Z$ further, gradually decreases the growth rate for tubes with large chiral angles, so that the highest growth rates occur for tubes with chirality close to $(2n,n)$, however at a rather low selectivity.\

Having discussed the influence of the energy of zigzag atoms, we now turn to the energy of other types of interface atoms. As barrier states always have one one-fold coordinated atom and $n+m-1$ two-fold coordinated atoms, all barrier energies depend on differences involving 1- and 2-coordinated atoms. Therefore, the effect of the energy of 1-coordinated atoms $E_X^1$ is opposite to that of 2-coordinated atoms $E_X$, where $X$ stands for $A$ or $Z$. In particular, a small $E_Z^1$ leads to a similar growth rate profile as obtained with a large $E_Z$ and vice versa. Similarly, the effect of armchair and zigzag energies on growth rates are opposite. Again, this can easily be understood from the barriers. What affects the growth rates is the relative energy of the different barriers. In particular, barriers 1, 3, 7, 9, 12, and 14 decrease with $E_Z$, while barriers 2, 4, 5, 6, 8, 10, 11, and 13 decrease with $E_A$. This means both $E_A$ and $E_Z$ change the relative height of the two groups of barriers, but in opposite directions. The types of growth rate maps obtained while varying these other interface energies ($E_Z^1$, $E_A$, and $E_A^1$) are not qualitatively different from those obtained while varying $E_Z$.\

\begin{figure}[htb!]
\begin{center}
\includegraphics{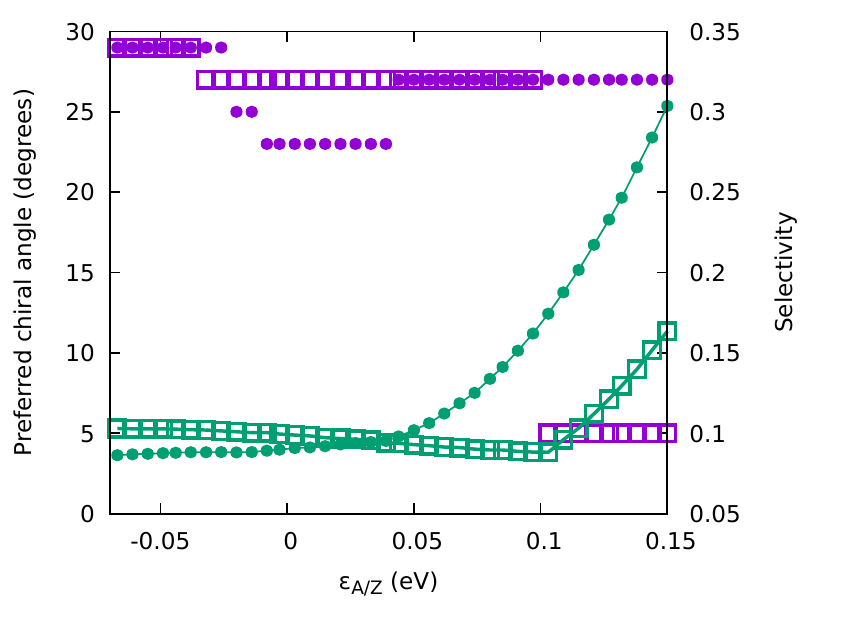}
\caption{\label{fig:ecaz}(Color online) Effect of ordering energy on favored chiral angle (left $y$-axis, purple symbols) and growth selectivity (right $y$-axis, green symbols). Full circles and open squares and correspond to $E_{A}$ = 0.05~eV and $E_{A}$ = 0.15~eV, respectively.}
\end{center}
\end{figure}

Finally, we consider the effect of the ordering energy parameter $\varepsilon_{A/Z}$ on the growth rate profiles. Figure~\ref{fig:ecaz} shows the chiral angle of the fastest growing tube (left $y$-axis) and the selectivity associated with this growth (right $y$-axis) as a function of $\varepsilon_{A/Z}$ for two different values of $E_A$. For negative values of $\varepsilon_{A/Z}$, growth profiles are qualitatively very similar to the baseline case, i.e. armchair tubes have the fastest growth, but with relatively low selectivity. For positive $\varepsilon_{A/Z}$, and $E_A=0.15$~eV, $(m,1)$ tubes have again the fastest growth, but for $E_A=0.05$~eV near armchair $(n, n-1)$ tubes grow the fastest with a higher selectivity. 
We are again dealing with a type of spiral growth, but dissimilar from the one encountered for the $(n,1)$ tubes, see Fig.~\ref{fig:growth_modes}(b). This growth process also occurs in a somewhat similar fashion for armchair tubes, however at each added new layer, a new kink needs to nucleate, which slows down the growth process. The barrier associated with this growth mode is barrier 3, with an effective height of 0.25~eV, which is much lower than the next barrier 4 at 0.45~eV. Other tubes grow much slower, because barrier 3 is not always available, which makes the time increments much larger.

\subsection{Examples of high selectivity}

\begin{figure}[htb!]
\begin{center}
\includegraphics{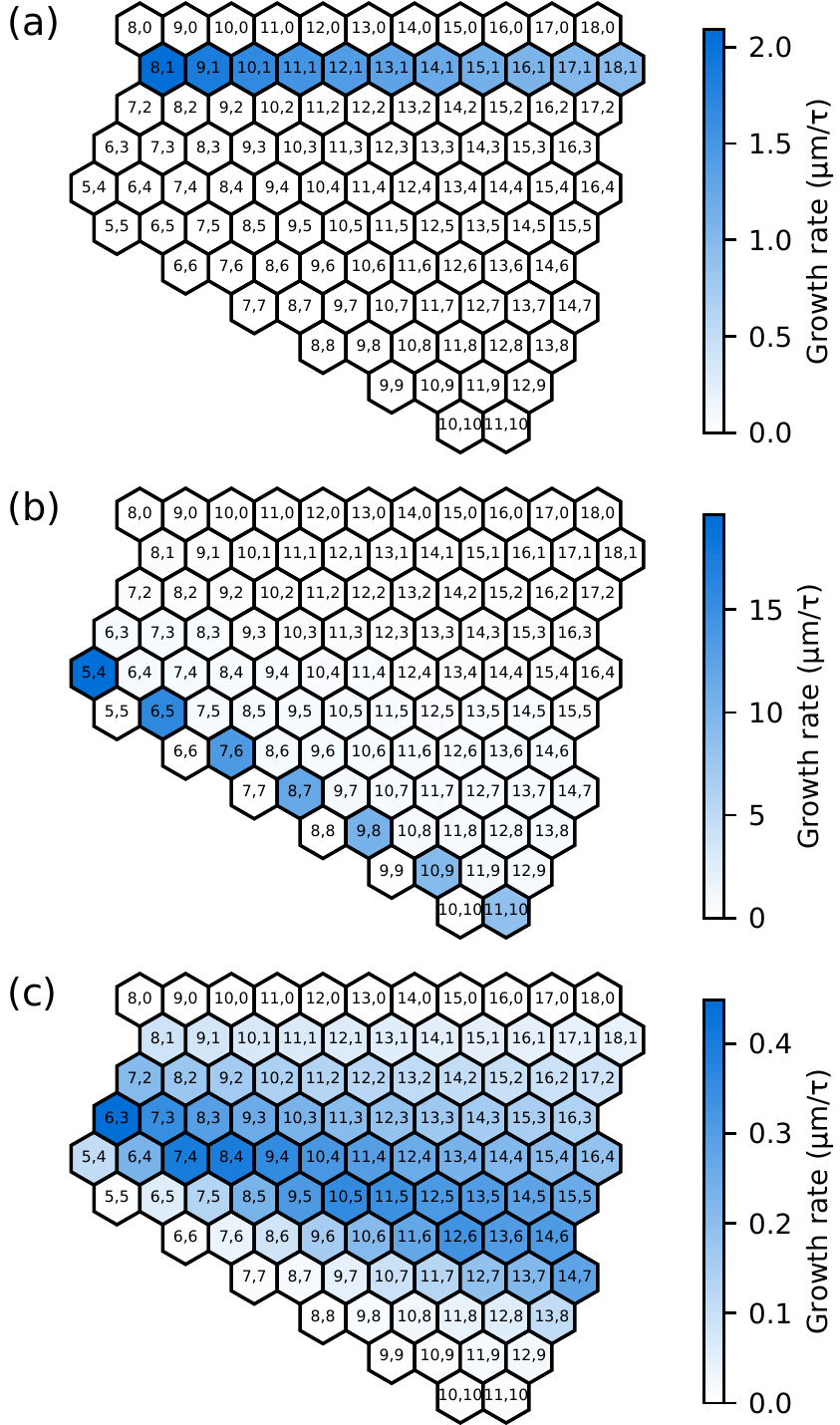}
\caption{\label{fig:maps}(Color online) Growth rates for three different types of chiral selectivity: (a) $m=1$, (b) $(n, n-1)$, and (c) $(2n, n)$.}
\end{center}
\end{figure}

Having understood how the parameters of the model affect growth rates, we explore the possibility to achieve a high selectivity on the chiral angle distributions. Figure~\ref{fig:maps} shows growth rate profiles for three cases of chiral angle selective growth. As before, all the following kMC runs are performed at 1000 K, however, the energy parameters are given a larger flexibility.

Panel (a) shows that a faster growth of $(n,1)$ tubes is obtained for $E_A = 0.172$~eV, $E_Z = 0.05$~eV, $E_A^1 = 0.8$~eV, $E_Z^1 = 1.5$~eV, $\varepsilon_{A/Z} = 0.1 $~eV, and $\Delta\mu_C = 0.1$~eV. There is only one possible edge structure for $(n,1)$ tubes, with $(n+m-2)$ zigzag and two armchair edge atoms, implying that spiral growth is their only possible mechanism. Since $(n,0)$ tubes do not grow, $(n,1)$ tubes are competing with tubes with larger chiral angles, which offer an increasingly large number of possible edges, by transforming armchair into zigzag edges and tilting the interface with respect to the tube axis. Because zigzag edges are more stable with these parameters, all $(n,m)$ tubes with $m> 1$ tend to maximize their number, while mostly ($60 \%$ of the configurations) keeping at least two armchair pairs instead of only one. The reason is probably the entropy gain that stabilizes the interface, while the two partial spiral developments hinder each other. If an anti-armchair motif (two neighboring zigzag edges with different inclinations) is present, which is often the case for $m>1$, an armchair pair can be added with barrier 5 and subsequently be removed using barrier 12. This leads to oscillations that pause growth over extended periods of time. The mechanism occurs only if barrier 4 (effective height 0.21~eV) is not available, because these barriers have an effective height of 0.41~eV and 0.85~eV which is much higher. These high barriers also make the time increment much larger, which slows growth even more. These are the reasons why all $(n,m)$ tubes with $m> 1$ have similar growth rates about one thousand times smaller than $(m,1)$ in the present conditions. We note however that $E_Z$ is generally larger than $E_A$ on metal catalysts~\cite{Ding2009, Liu2010}, explaining why this calculated preference $(n,1)$ tubes is rarely obtained.\\

By setting $E_A = 0.1$~eV, $E_Z = 0.4$~eV, $E_A^1 = 0.72$~eV, $E_Z^1 = 0.92$~eV, $\varepsilon_{A/Z} = 0.2$~eV and $\Delta\mu_C=0.1$~eV a highly selective growth of $(n,n-1)$ tubes is achieved with a decent yield, as shown in panel (b). With these parameters, the lowest effective barriers are 3 and 7 at only $0.02$~eV. The next higher ones are at $0.22$~eV and $0.42$~eV, which makes the growth very selective. The low energy of the armchair edges and the strongly repulsive ordering energy stabilize edges with $(n+m-2)$ armchair and two zigzag atoms, which leads to an efficient spiral growth, with a high yield. Here, armchair tubes do grow slowly because parameters prevent forming oblique interfaces ($E_Z>> E_A$). In addition, nucleating a new ring, that requires the formation of $\bm{r} = 4$ A/Z contacts associated with a high barrier due to large positive $\varepsilon_{A/Z}$, needs a lot of time which reduces overall growth rates. Practically, such harsh conditions are difficult to achieve in experiments, and a fraction of metallic armchair tubes are usually produced at the same time as the semi-conducting $(n,n-1)$ tubes. It would otherwise lead to a much coveted growth of semi-conducting tubes!

The parameter range to selectively grow tubes with intermediate chiral angles, close to $(2n, n)$ chiralities is much narrower, and the resulting selectivity is smaller. Panel (c) presents a map obtained for $E_A = 0.16$~eV, $E_Z = 0.25$~eV, $E_A^1 = 0.35$~eV, $E_Z^1 = 1.20$~eV, $\varepsilon_{A/Z} = -0.09$~eV and $\Delta\mu_C = 0.03$~eV. In this case barriers 2 (effective height $0.07$~eV) and 12 (effective height $0.13$~eV) are significantly lower than the others and control the growth sequence: dimers are added between two consecutive armchair pairs, with significant fluctuations. The growth process of the $(8, 4)$ tube, which is one of the fastest, is shown in Fig.~\ref{fig:growth_modes}(c) and in the corresponding video. This mechanism is rather inefficient and the yield is low. In Fig.~\ref{fig:maps}(b) and (c), we can see that the maximum growth rate of $(2n,n)$ tubes is about $40$ times lower than the maximum growth rate of $(n, n-1)$ tubes, each tube family under its optimal growth condition. Experimentally, an extremely low yield has been reported for the growth of $(12,6)$ tubes using Co-W bimetallic catalysts~\cite{An2016}, while recent investigations suggest that this low yield could be compensated by a longer catalyst lifetime~\cite{Zhang2020}. The former reports a selectivity around $50-70\%$, while the  abundance of $(12,6)$ reached approximately $22\%$ among the diameter range of $0.81 - 1.53$ nm in the latter study.   Our results are clearly compatible with these experimental data. We can thus speculate that the $(2n,n)$ selective growth reported in Refs.~\onlinecite{Yang2014, Zhang2017} also proceeds through this mechanism, even though no information is given on the yield and the reported selectivity is much larger.

\section{Discussion}

A central assumption in this study is to consider ideally crystalline tube structures, without defects. Once created during the growth process~\cite{Hisama2018}, defects can be healed in the vicinity of the catalyst~\cite{Karoui2010} or remain present. Whether defect healing is correlated with the tube structure~\cite{Page2010b, Diarra2012} is still an open question, and we assume that the presence of defects has no influence, other than detrimental, on the mechanisms fixing the chirality during growth. Since our goal is to answer the question if growth kinetics can lead to a significant chiral selectivity, we have to consider ideal crystalline tube  structures and compare their growth rates: ideal tube structures are formed with carbon hexagons only, and we consequently consider only Monte Carlo moves that add or remove carbon dimers at the edge of the tube. A consequence is the somewhat oversimplified definition of the barrier states. On the upside, this avoids discussing detailed atomistic mechanisms that remain inaccessible experimentally, and the energies of one-fold coordinated species at the edge $E_Z^1$ and $E_A^1$ can be seen as variables controlling barrier heights that are necessarily present in the process, though essentially unknown, but of prime importance on the kinetics. 

Our simulations show that SWNT growth is determined by complex processes involving the relative stability of armchair and zigzag edges, the degrees of freedom of the edge structure, and the associated configurational entropy, as well as the energy barriers controlling the incorporation of incoming carbon atoms. The entanglement of these contributions makes it difficult to develop a rational approach to experimentally achieve a high selectivity, but we expect this contribution to promote it. The spiral growth process envisioned by Ding \textit{et al.}~\cite{Ding2009} is indeed sometimes observed. However, the conclusions were drawn too hastily. In fact, not only $(n,n-1)$ but also $(n,1)$ series may grow fast through this process, but under some conditions more disordered growth, involving a larger set of carbon incorporation barriers proves more efficient than pure spiral growth. 

In our previous modeling~\cite{Magnin2018}, we assumed that the interface was perpendicular to the tube axis, with fixed numbers of armchair and zigzag edge atoms. Since oblique interfaces have been observed experimentally, the natural step forward is to assume that edges simply have $n+m$ edge atoms. Allowing for even longer edges would lead to unphysically large interface meandering. In the present model, zigzag tubes cannot nucleate and grow. DFT calculations by Ding \textit{et al.}~\cite{Ding2009} show that the probability of nucleating zigzag tubes is very low. Another consequence is that the configurational entropy of the edge is larger on the armchair side. A full development of the thermodynamic aspects of these assumptions should be performed to obtain new $(n,m)$-dependent probability distributions, that, combined with the growth rates calculated here would produce narrower diameter dependent distributions. Maps showing the abundance of each tube chirality could then be more significantly compared with experimental data than the purely kinetic data presented here. Such calculations are beyond the scope of this introductory paper.\

\section{Conclusions}

We have developed a simple yet powerful kinetic Monte Carlo modeling of SWNT growth that allows comparison of growth rates of large sets of tubes with various chiralities. Significant differences in growth rates can be expected by tuning the interfacial energies. This growth selectivity of kinetic origin complements the thermodynamic selectivity~\citep{Magnin2018} operating at the nucleation stage. In principle, interfacial energies depend on the growth mode~\cite{He2018} and the adhesion energy of the tube to the catalyst, which can be evaluated by DFT calculations~\cite{Ding2008, Hedman2019, Qiu2019, Chao2020}. In practice, the interfacial properties should be adjusted by an adequate choice of  catalyst and experimental conditions. Direct validation of our results is difficult due to the scarcity of measurements of  growth rates of individual tubes~\cite{Rao2012, Koyano2019}. However, the progress currently made in this direction~\cite{Jourdain2020} is a strong motivation for this work.

Beyond the general analysis of the growth kinetics which critically depends on the identification of all possible barriers, we have identified growth parameters (interfacial energies, carbon chemical potential, temperature) leading  to experimentally observed selectivity patterns. Baseline parameters for which all insertion and all etching barriers are equal lead to a slight selectivity towards large chiral angles, commonly observed for tubes grown with Fe in perpendicular mode~\citep{He2018}. Sharper near armchair selectivity is obtained for interfacial energy parameters $E_A < E_Z$ and when energy barriers $3$ and $7$ are the weakest. The first condition corresponds to $(E_A^{adh}+E_A^{db}) < (E_Z^{adh}+E_Z^{db})$, which is observed for many catalysts~\cite{Ding2008}. In such a case, the difference $E_A^{adh}-E_Z^{adh}$ between the adhesion energies of armchair and zigzag edges is less than the difference in dangling bond energies $E_Z^{db}-E_A^{db} \approx 1.1 eV$. Finally, we have identified conditions for growing tubes with a chiral angle around $19^\circ$, which come at the price of a low yield, as observed experimentally, because of a rather inefficient growth mechanism.

Our simulations also highlight the importance of including both insertion and removal mechanisms in the kMC algorithm. With the non-selective baseline parameters, growth and etching are symmetrical for all tube chiralities, as observed experimentally~\cite{Koyano2019}. However, our calculations suggest that this is not a general feature though. Thus, taking advantage of conditions in which the $(n,m)$ dependence of the growth and etching processes is asymmetric might open new opportunities for selective growth. 

\begin{acknowledgments}
Support from the French research funding agency (ANR), under grant 18-CE09-0014-01 (GIANT) is gratefully acknowledged. The authors thank Drs H. Amara and F. Ducastelle for stimulating discussions.
\end{acknowledgments}

\section*{Data availability}
The data that support the findings of this study are available from the corresponding author upon reasonable request.

\appendix

\section{Depiction of all 14 possible kMC barriers}

\begin{figure*}[htb!]
\begin{center}
\includegraphics{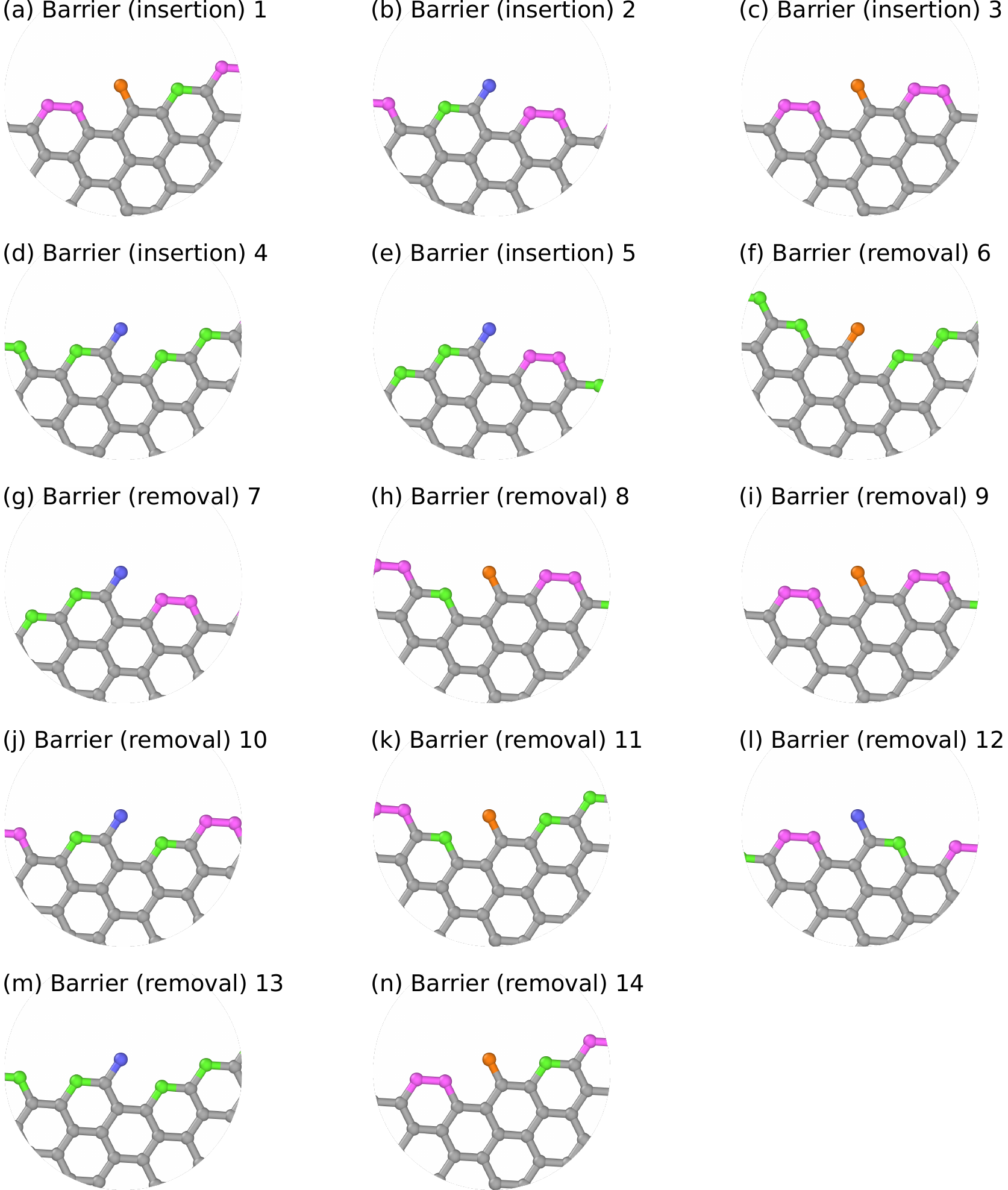}
\caption{\label{fig:barriers}(Color online) (a--n) Barrier states (1--14) encountered in the kMC simulations. The barrier states always contain a mono-coordinated atom (either attached to an armchair atom [blue sphere], or attached to a zigzag atom [orange sphere]) at the interface with the catalyst. Zigzag and armchair atoms are shown as green and violet spheres, respectively.}
\end{center}
\end{figure*}

Fig.~\ref{fig:barriers} shows a graphical representation of all 14 barrier states listed in table~\ref{tab:barriers}. Note that seemingly similar configurations may differ by the number $\bm r$ of armchair/zigzag contacts.

\section{\label{sec:si}Derivation of eq.~(\ref{eq:g_tot})}

In the ``baseline case'' all barriers $\Delta E_i=\Delta E$ are equal. Therefore, there are only two different rates at play [compare eq.~(\ref{eq:defRate})], one for insertion and one for removal pathways. We designate the number of insertion or removal sites $N$. It turns out that there are always the same number of insertion and removal pathways. This number is equal to the number of armchair-type C-atoms, and it depends thus on the configuration of the CNT edge at each kMC step. The total rates for insertion and removal are therefore
\[
k_{\text{tot}}^{\text{ins}} = N \nu_0 \: \text{exp}\left(- \Delta E /k_BT\right) \text{exp}\left( \Delta\mu_C /k_BT\right)
\]
and
\[
k_{\text{tot}}^{\text{rmv}} = N \nu_0 \: \text{exp}\left(- \Delta E /k_BT\right) \text{exp}\left(-\Delta\mu_C /k_BT\right).
\]
The sum of all rate constants, $ k_{\text{tot}} = k_{\text{tot}}^{\text{ins}}+k_{\text{tot}}^{\text{rmv}} $ is required for the calculation of the time increment. It corresponds to a random number drawn from the distribution $p\left(t\right) = k_{\text{tot}}\text{exp}\left(-k_{\text{tot}}t\right),$ which leads to an average time increment of $\Delta t_{\text{avg}} = 1/k_{\text{tot}} .$
In order to compute growth rates, we also calculate the number of atoms added at each kMC step. Removal and insertion events are selected randomly, but proportional to their respective rate constants. Considering that at each step two C-atoms are either added or removed, we are left with
\[
\Delta N_{\text{avg}} = 2k_{\text{tot}}^{\text{ins}}/k_{\text{tot}} -2 k_{\text{tot}}^{\text{rmv}}/k_{\text{tot}}
\]
on average. The growth rate in terms of atoms per unit of time is then:
\begin{eqnarray*}
G_{\text{avg}} =\Delta N_{\text{avg}}/\Delta t_{\text{avg}} \\
= 4 \times N \nu_0 \: \text{exp}\left(- \Delta E /k_BT\right) \text{sinh}\left( \Delta\mu_C /k_BT\right).
\end{eqnarray*}
As a side note, the unit of the prefactor can be converted from atoms per $\tau$ to nm/$\tau$. This is done by multiplying the prefactor with the length $l= \left(\varrho c \right)^{-1}$ that one atom adds to the CNT. The areal number density of graphene $\varrho = 4/(3 \sqrt{3} d_{\text{CC}}^2) \approx 37.12 \; \text{atoms/nm}^2$ assuming the nearest neighbor distance $d_{\text{CC}} = 0.144\; \text{nm}$, and $c= \sqrt{3} d_{\text{CC}} \times \sqrt{n^2+nm+m^2}$ the circumference of the tube. This factor obviously depends on the chiral indices $n$ and $m$.

\section{\label{sec:fluctuation}Fluctuation-dissipation theorem}
\begin{figure}[htb!]
\begin{center}
\includegraphics{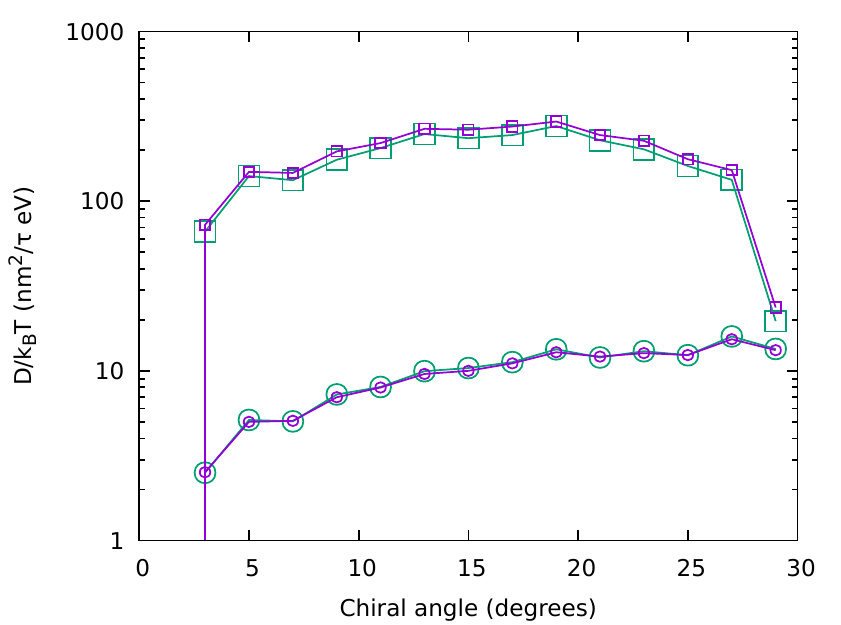}
\caption{\label{fig:fluctuations}(Color online) The diffusivity divided by the thermal energy, $D/{k_BT}$, obtained at $\Delta \mu_C=0$ (large, green symbols) and the mobility $v(\Delta \mu_C)/(\Delta \mu_C/l)$ in the limit $\Delta \mu_C\to0$ (small, purple symbols), which are predicted to be equal by the fluctuation-dissipation theorem, eq.~(\ref{eq:fdt}). Circles and squares correspond to kMC simulations using the ``baseline'' parameter set and $E_Z=0.5$~eV, respectively.}
\end{center}
\end{figure}

In equilibrium, the height of the interface undergoes 1-D diffusion, with a diffusion constant
\[
D=\frac{1}{2}\lim_{t \rightarrow \infty} \frac{\langle [h(t_0+t)-h(t_0)]^2 \rangle}{t},
\]
whilst for finite $\Delta\mu_C\neq0$ the interface drifts with a velocity
\[
v(\Delta\mu_C) = \lim_{t \rightarrow \infty} \frac{\langle h(t_0+t)-h(t_0) \rangle}{t}.
\]
In this setting, with an effective energy gradient $\Delta\mu_C/l$, the fluctuation dissipation theorem reads\cite{Coffey2012}
\begin{equation}\label{eq:fdt}
 \lim_{\Delta\mu_C \rightarrow 0} \frac{v(\Delta\mu_C)}{(\Delta\mu_C/l)} = \frac{D}{k_BT}.
\end{equation}
In Fig.~\ref{fig:fluctuations} we show this relation to be satisfied across the entire range of chiralities with two sets of energy parameters, thus validating the thermodynamic consistency of the present kMC approach.

\newpage
\nocite{*}
\bibliography{kmc_cnt_growth}

\end{document}